\documentclass[twocolumn,floatfix,showpacs,aps,superscriptaddress,letterpaper]{revtex4}

\usepackage{times}
\usepackage{mathptmx}

\usepackage{dcolumn}
\usepackage{graphicx}
\usepackage{mathrsfs}
\usepackage{amsfonts}
\usepackage{amsbsy}
\usepackage{url}
\usepackage{latexsym}
\usepackage{verbatim}
\usepackage{amsmath}
\usepackage{hyperref}

\begin{document}

\newcommand{\ud}{\mathrm{d}}

\newcommand{\tfull}{\mathbf{t}}
\newcommand{\thatfull}{\hat{\mathbf{t}}}
\newcommand{\rfull}{\mathbf{r}}
\newcommand{\rp}{\mathbf{r}^{\bot}}
\newcommand{\tp}{\mathbf{t}^{\bot}}
\newcommand{\rz}{\mathbf{r}^{\parallel}}
\newcommand{\tz}{\mathbf{t}^{\parallel}}
\newcommand{\ps}{\partial_s}
\newcommand{\kp}{\mathbf{v}}

\newcommand{\kT}{k_{\mathrm{B}}T}
\newcommand{\lp}{l_{p}}
\newcommand{\fb}{f_{b}}
\newcommand{\lb}{l_{b}}
\newcommand{\Nfil}{N_{\mathrm{fil}}}

\title{The limits of filopodium stability}
\author{Sander Pronk}
\affiliation{Department of Bioengineering, U.C. Berkeley, Berkeley, CA 94720, USA}
\author{Phillip L. Geissler}
\affiliation{Department of Chemistry, U.C. Berkeley, Berkeley, CA 94720, USA}
\author{Daniel A. Fletcher}
\email{fletch@berkeley.edu}
\affiliation{Department of Bioengineering, U.C. Berkeley, Berkeley, CA 94720, USA}

\begin{abstract}
Filopodia are long, finger-like membrane tubes supported by cytoskeletal filaments. Their shape is determined by the stiffness of the actin filament bundles found inside them and by the interplay between the surface tension and bending rigidity of the membrane. Although one might expect the Euler buckling instability to limit the length of filopodia, we show through simple energetic considerations that this is in general not the case. By further analyzing the statics of filaments inside membrane tubes, and through computer simulations that capture membrane and filament fluctuations, we show under which conditions filopodia of arbitrary lengths are stable. We discuss several \emph{in vitro} experiments where this kind of stability has already been observed. Furthermore, we predict that the filaments in long, stable filopodia adopt a helical shape. 
\end{abstract}

\pacs{87.16.Qp, 87.16.af, 87.17.Pq}

\maketitle

\section{Introduction}

Filopodia are slender protrusions from a cell's exterior surface, which may act as mechano-sensors during axon growth and cell movement\cite{Alberts2002,Wood2002,Gupton2007}. Their shapes and stability are determined by a mechanical interplay between the bounding lipid membrane and enclosed bundles of the filamentous protein actin. Tension and bending rigidity of the membrane resist formation and growth of filopodia, while actin filaments, running parallel to the long axis of the filopodium and rooted in the cytoskeleton, provide the counterbalancing force against membrane retraction.

The tubular shape of membrane extensions, usually called membrane tethers or tubes in the absence of a filament bundle, reflects a compromise between energetic costs of stretching and bending the membrane. At a certain tube radius $R$, the reward for reducing surface energy (which scales as $R$) precisely balances the concomitant penalty for increasing curvature (which scales as $1/R$). Resulting from this balance is a membrane energy that grows linearly with the tube's length $L$, giving rise to a 
longitudinal restoring force\cite{Mogilner2005, Der'enyi2002,Koster2005}.

By itself, a bundle of actin filaments should behave under compression much like a simple elastic rod. Compressive forces below a certain threshold $f_\mathrm{b}$ induce little deformation. Beyond that threshold the rod becomes extremely pliable, undergoing a long-wavelength instability known as Euler buckling. Because $f_\mathrm{b}$ decreases quadratically with a rod's length, a growing actin bundle under fixed load is expected to buckle and collapse at a critical length, $l_\mathrm{b}$.

Together, these arguments would seem to imply an upper limit on filopodial growth: once the length of a filopodium exceeds the Euler buckling length, the filament bundle can no longer sustain the restoring force of the membrane tube, leading to collapse. Calculations based on this notion suggest a limiting length of $1-2 \mathrm{\mu m}$\cite{Atilgan2006, Mogilner2005}. By contrast, filopodia several tens of $\mathrm{\mu m}$ in length have been observed in experiment\cite{Wood2002,Gustafson1961}. Stability of long filopodia has been rationalized as a consequence of tight bundling of actin filaments such as done by the protein fascin. A quadratic increase of bundle stiffness with the number of tightly linked filaments, however, is insufficient to explain the observation of filopodia over $10 - 20 \mathrm{\mu m}$ in length.

In this article we reconsider the buckling of a semi-flexible filament bundle inside a membrane tube, paying careful attention to the compatibility of membrane and bundle geometries. In our calculations the buckling instability is removed by the constraint that the tube must enclose filaments as they deform. Contrary to conventional pictures, presence of an enclosing membrane \emph{stabilizes} the bundle against buckling, rather than causing it, so that filaments in a sufficiently thin tube may grow to arbitrary length without collapse. We present Monte Carlo simulations of a worm-like chain inside an elastic tube, fully incorporating effects of thermal fluctuations, which verify this surprising stability.

\section{The energetics of buckling filopodia}

We describe the conformation of a semi-flexible filament (or bundle of filaments) by a parameterized curve $\rfull(s)$ with inextensibility condition $\left| \partial \rfull(s) / \partial s \right| = 1$. The corresponding energy is that of a worm-like chain,
\begin{equation}
E_{\mathrm{fil}}[\rfull(s)]  = \frac{\lp \kT}{2} \int_0^l \ud s 
\left| \frac{\partial^2 \rfull(s)}{\partial s^2}
\right|^2,
\label{Esemi}
\end{equation}
where $l$ is the bundle's contour length, and $\lp$ is its  effective persistence length: throughout this text we will assume uncross-linked bundles with an effective persistence length $\lp = \Nfil \lp^*$ where $\Nfil$ is the number of filaments in the bundle and $\lp^*\approx 15\ \mathrm{\mu m}$ for a single actin filament.

This model yields an Euler buckling force $\fb= \lp \kT \pi^2/4 l^2$ at which sinusoidal deformations of period $2l$ become favorable at all amplitudes. For a growing rod under fixed compressive load $f$, the buckling length is therefore $\lb=\sqrt{\lp \kT \pi^2/4 f}$.  The energy of the membrane tube includes contributions from both surface tension and bending energy. For a cylindrical geometry the standard model of Helfrich\cite{Gompper1997} yields\cite{Der'enyi2002}
\begin{equation}
E_{\mathrm{tube}} = \left(\frac{\kappa \pi}{R} +  2 \pi \gamma R \right) L 
\equiv f_{\mathrm{tube}} L,
\label{Ftube}
\end{equation}
where $\gamma$ is the surface tension, $\kappa$ is the bending rigidity, $L$ is the tube's length, and $R$ is its radius. Because the energy grows linearly with $L$, a constant force $f_{\mathrm{tube}}$ acts longitudinally against the overall filopodium contour length\footnote{At longer tube lengths the surface tension increases as the thermal undulations of the vesicle are pulled out\cite{Koster2005}, which actually turns out to stabilize the filopodia further against collapse. This effect may, however, provide a maximum length where the force equals the stall force of the growing filament bundle.}.

\begin{figure}
\center
\includegraphics[width=0.90\linewidth]{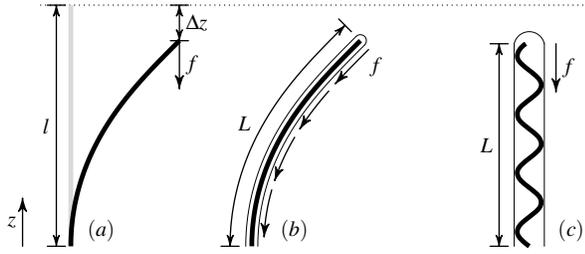}
\caption{Absence of Euler buckling in narrow filopodia. If an initially straight elastic rod ($a$) experiences a sufficiently large downward force $f$, it will buckle, because the energy gain $f\Delta z$ exceeds the bending energy of the rod. By contrast, in a filopodium ($b$) where the compressive force exerted by the membrane is directed along the contour of the supporting filament bundle, denying it the potential energy gain that leads to buckling. 
A filopodium may lower its energy, however, by adopting a helical configuration ($c$). Here, the energy decreases due to shortening of the membrane tube.
 }

\label{fig:bend}
\end{figure}

\subsection{A simple argument against Euler buckling}

Fixed compressive force on a rod, i.e., an external potential that decreases in proportion to the rod's end-to-end distance, is a crucial ingredient of the Euler buckling scenario. As described above, a {\em straight} membrane tube exerts such a longitudinal force. 

A buckling filament bundle, however, does not remain straight. In a narrow filopodium the membrane will accommodate the filaments' deflection by deforming congruently, as depicted in Fig.~\ref{fig:bend}($a$) and ($b$). As a result the compressive forces tending to retract the tube will follow the contour of the bundle, no longer directed along the rod's end-to-end distance. 

If the contour length of the tube does not decrease, deflecting the bundle begets no energetic reward. In the limit of a vanishingly thin filopodium, the geometric constraint of enclosure will thus negate any energetic gain from retracting the filament, preventing enclosed actin filaments from buckling.

In other words: since any bending of the filament will lead to bending of the membrane, the compressive force $f_{\mathrm{tube}}$ that acts to shorten the membrane tube will be exerted along the contour of the filament. Though ``compressive'', this force will in effect counteract buckling by adding membrane bending energy to the filament bending energy.

\subsection{Buckling of filopodia of finite radius}
\label{sec:finrad}

The radius of an empty membrane tube can be estimated from Eq.~\ref{Ftube}.  As a function of $R$ this energy is minimal at
\begin{equation}
R = \sqrt{\frac{\kappa}{2\gamma}}.
\label{tuber_equil}
\end{equation}
For typical values of membrane rigidity $\kappa \approx 40 \ \kT$ and surface tension $\gamma \approx 0.0025 \ \kT/\mathrm{nm^2}$, Eq.~\ref{tuber_equil} gives $R = 89 \ \mathrm{nm}$.  (For cell membranes, values for $\kappa$ range from $20-80\ \kT$, and $\gamma$ ranges from $0.0013-0.25 \ \kT/\mathrm{nm^2}$\cite{Henriksen2004,Upadhyaya2004,Olbrich2000,Evans1983,Titushkin2006}.) 
    
Adding Eqs.~\ref{Esemi} and \ref{Ftube} gives the total energy of a semi-flexible filament enclosed by a membrane tube: 
\begin{equation}
E = \frac{\lp}{2} \int_0^l \ud s 
\left( 
\frac{\partial^2 \rfull(s)}{\partial s^2}
\right)^2
+ \left( \frac{\pi \kappa}{R} + 2 \pi \gamma R \right) L.
\label{Fsemitube}
\end{equation}
Our analysis of filopodium buckling will focus on minimizing Eq.~\ref{Fsemitube} with respect to $L$, $R$, and ${\bf r}(s)$, subject to the constraint of enclosure. The global minimum will always correspond to a tube of zero length and infinite radius ``enclosing'' a straight filament lying parallel to the flat membrane, i.e., complete collapse. Our arguments above suggest, however, that this configuration may be very difficult to reach. Beginning from an initial state of narrow protrusion, collapse would require that large energy barriers be surmounted through costly bending fluctuations. If other local energy minima exist, and can be accessed with modest deformation, they are likely to be very stable.


Any plausible mode of deformation would maintain the bundle's contour within a small radius, as could be accomplished by a helical configuration (see Fig.~\ref{fig:bend}($c$)). Below, we consider in detail the energetics of a helical bundle circumscribed by a cylindrical membrane tube. While this choice is not unique, it does allow for efficient reduction of the tube's length without widening or bending the cylinder. This scenario, which we refer to as ``helical buckling'', is described mathematically by
\begin{equation}
\rfull(s)=\left(\begin{array}{c}
	R \cos 2 \pi n s \\
	R \sin 2 \pi n s \\
	s \sqrt{1-n^2 \pi^2 R^2}
    \end{array}\right),
\label{helixrdef}
\end{equation}
where $n$ is the number of helix windings per unit contour length.  Eq.~\ref{helixrdef} ensures filament in-extensibility as well as enclosure within a membrane tube of radius $R$ and length $L = l\sqrt{1-n^2 \pi^2 R^2}$.  Notice that this parameterization includes as limiting configurations both an undeformed filopodium ($L/l=1$) and a completely collapsed filopodium ($L/l=0$). 

\begin{figure}[t]
\centering
\includegraphics[width=0.48\linewidth]{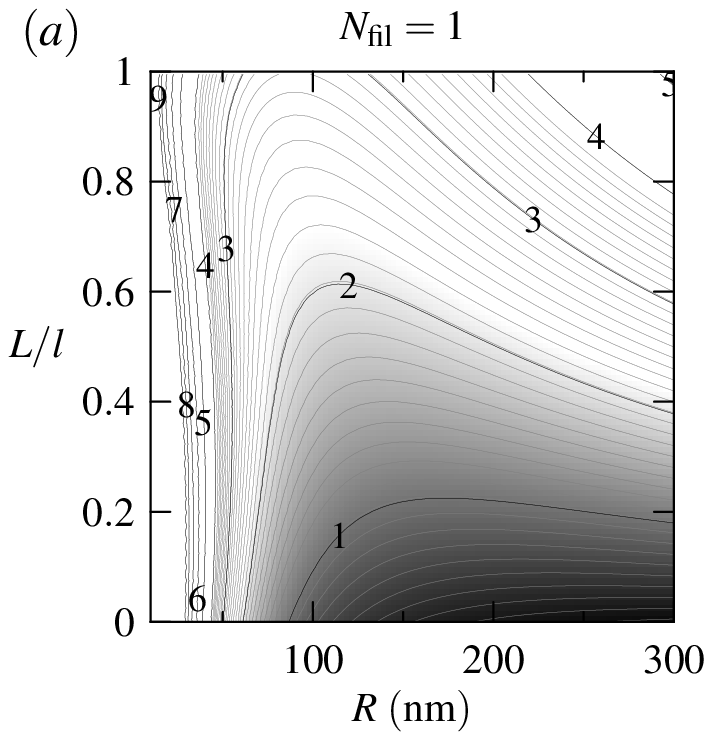}
\includegraphics[width=0.48\linewidth]{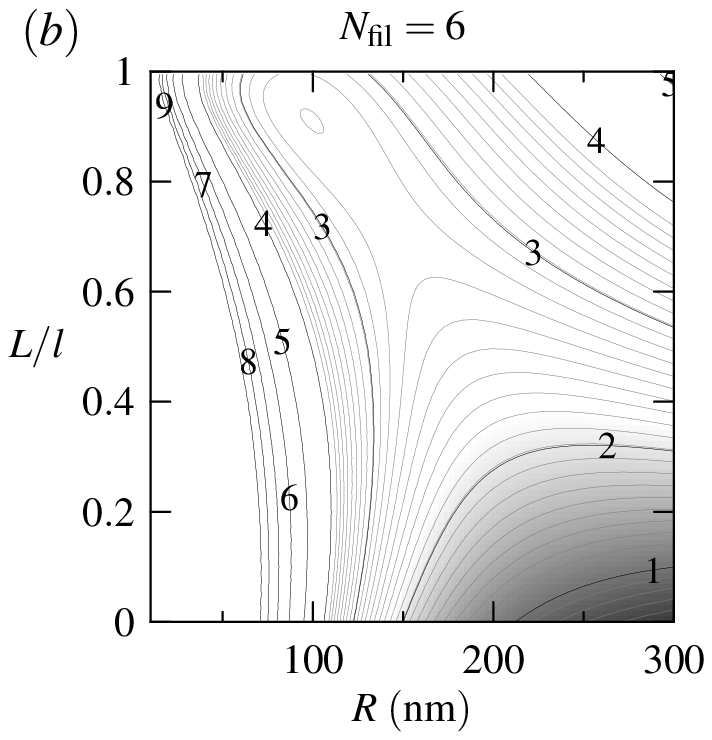}
\caption{
    Contour plot of the combined membrane and filament energy per unit contour length from Eq.~\ref{Esemitubehelical}, as a function of tube radius $R$ in nanometers, and of the ratio of the tube length $L$ and the filament contour length $l$ (a measure of helicity). The membrane bending rigidity is $\kappa=40\ \kT$, and its surface tension is $\gamma=0.0025 \ \kT/\mathrm{nm^2}$.  The plot ($a$) shows the energy for 1 filament, and plot ($b$) is for 6 filaments. Darker shades stand for lower energies; numbers label contours in units of $\kT/\mathrm{nm}$. Note the presence of a local energy minimum in ($b$). 
}
\label{fig:helicalcontour}
\end{figure}

Fig.~\ref{fig:helicalcontour} shows the energy per contour length of a helically buckled filopodium,
\begin{equation}
\frac{E}{l} =	\frac{\lp}{2}\frac{1}{R^{2}}
	    \left( 1 - \frac{L^2}{l^2}\right)^2
	    + 
	    \left( 
		    \frac{\pi \kappa}{R} + 2 \pi \gamma R
	    \right) \frac{L}{l}
    \label{Esemitubehelical}
\end{equation}
as a function of $R$ and $L/l$. For the values of $\kappa$ and $\gamma$ considered, a bundle comprising of just one filament ($\Nfil=1$) possesses a single energy minimum at $L/l=0$, i.e., it is unstable to collapse. But stability against collapse can be achieved with a modest increase in the number of filaments. With only six filaments a local energy minimun appears at approximately one winding per $1500\ \mathrm{nm}$ ($R\approx 100 \ \mathrm{nm}$, $L/l \approx 0.9$). Not surprisingly, the corresponding radius slightly exceeds that of an empty tube, reflecting the radial force generated by this mode of bundle deformation. The energy barrier to collapse in this case, roughly $0.1\ \kT/\mathrm{nm}$, is indeed substantial for a filopodium more than $100 \mathrm{nm}$ in length.\footnote{The filopodia we consider are well past the stage of initiation; filament bundle anchoring boundary conditions critical for initiation are therefore assumed to play a minor role. For a more complete treatment of filopodium initiation, see for example Refs~\cite{Der'enyi2002} and \cite{Liu}}

\begin{figure}[t]
\centering
\includegraphics[width=0.75\linewidth]{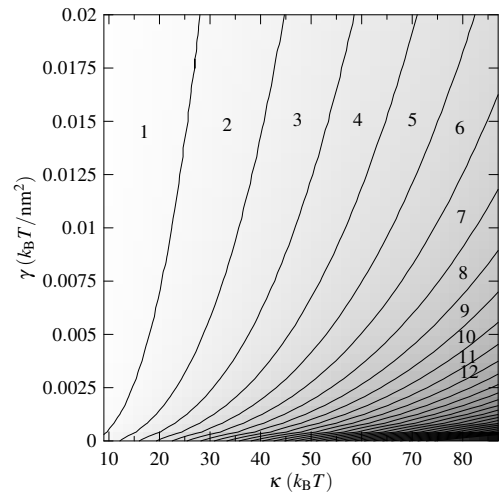}
\caption{
    Contour plot of the number of filaments with $\lp^*=15 \mathrm{\mu m}$ required for a stable, helically buckled filopodium as a function of curvature rigidity $\kappa$ and surface tension $\gamma$. Increasing the surface tension $\gamma$ yields narrower membrane tubes, stabilizing the filopodium against buckling.}
\label{fig:min_N_helical}
\end{figure}

By the same reasoning, it is possible to find a minimum number of filaments that keeps the filopodium stable for any combination of membrane surface tension and bending rigidity. In Fig.~\ref{fig:min_N_helical}, we plot this number against plausible values for $\kappa$ and $\gamma$, and see that it is less than 10 for most of this range, implying that a small number of filaments is enough to stabilize arbitrarily long filopodia against buckling. This number is well within the range of what is commonly thought to be the actual number of filaments in filopodia\cite{Mogilner2005} and is similar to the number of filaments thought to be required to nucleate a filopodium\cite{Atilgan2006,Liu}.

If the filopodium as a whole experiences an external force, e.g. when it is pushing against an obstacle, it will generally not be stable against Euler buckling. This situation arises in experiments such as those performed by Liu et. al.~\cite{Liu}, where filopodium-like protrusions grow into the lumen of a vesicle and contact the other end of the vesicle, and buckle. 

Another experimental observation of stability --- and instability --- against buckling is found in Ref.~\cite{Kuchnir1997}. There, a microtubule is grown inside a vesicle in such a way that it forms membrane-enclosed protrusions on both ends of the vesicle. Because the membrane envelops these protrusions, the microtubule is stable against buckling wherever it is in a protrusion. Inside the vesicle, however, the microtubule is not enveloped by a membrane tube and thus proceeds to buckle.

\section{Simulations}

In order to check whether the stability argument described in the previous section is valid, we performed Monte Carlo simulations of a semi-flexible rod in a membrane tube. In these simulations, the membrane is modeled as a triangulated sheet with dynamic re-triangulation as described in~\cite{Ho1990,Gompper1997}, with bending energy calculated as in Ref. \cite{Gompper1996}. The rod, which represents the filament bundle, is discretized into many sections of constant length. 

Both the membrane vertices, and the filament discretization points consist of excluded volume enforcing the impenetrability of the membrane to the filament and to itself. The effective bending rigidity, which is influenced by the presence of exclusion spheres on the membrane triangulation vertices, is measured and re-calibrated by measuring the radius that an empty membrane tube adopts. 

\begin{figure}[tb]
\centering
\includegraphics[width=1.00\linewidth]{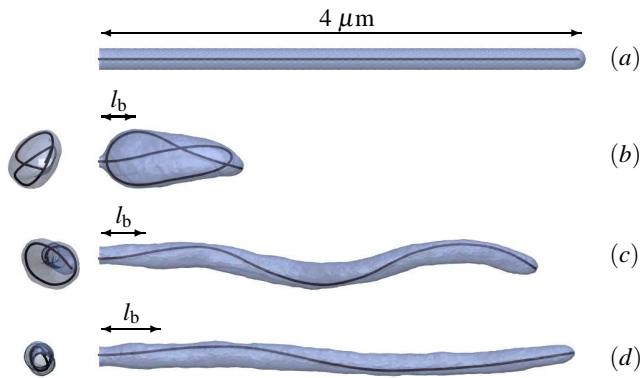}
\setlength{\unitlength}{\linewidth}
\caption{Simulation snapshots showing the membrane and filament bundle inside it (perpendicular to, and along the longitudinal axis). Filopodia with $4\  \mathrm{\mu m}$ contour length begin each simulation in an elongated configuration shown in $a$. The filament bundles with stiffnesses corresponding to 6 (for $b$), 10 (for $c$), and 18 (for $d$) filaments (with parameters similar to those in Fig.~\ref{fig:helicalcontour}), are shown after approximately $8\cdot10^6$ MC steps per membrane triangle vertex. The Euler buckling lengths $l_\mathrm{b}$ for these configurations are $0.28\  \mathrm{\mu m}$ ($b$), $0.36\  \mathrm{\mu m}$ ($c$) and $0.49\  \mathrm{\mu m}$ ($d$) respectively. }
\label{fig:simmem}
\end{figure}

The initial geometry shown in Fig.~\ref{fig:simmem} is a straight filopodium with a spherical cap and a straight filament of $4\  \mathrm{\mu m}$, which exceeds the Euler buckling length in all cases. The 4270 triangles of the membrane are distributed over the surface so that that they are close to equilateral. 

The membrane vertices and filament points are free to move except at the `open' end, where the filament bundle is held at fixed orientation and the end-vertices of the membrane are restricted to the plane perpendicular to the initial filament direction. 

As shown in Fig.~\ref{fig:simmem}, simulated filopodia are stable at lengths far beyond their Euler buckling length, even though the membrane deviates noticeably from a straight cylindrical tube (if there were no membrane curvature energy, the membrane would adopt a helicoidal shape\cite{Catalan1842}). This membrane deformation almost disappears as the filament bundle is made stiffer.

We do, however, find that the simulated filopodia are only stable against buckling at higher filament stiffnesses (larger numbers of filaments) than the analysis leading to Fig.~\ref{fig:min_N_helical} would predict. For the parameters used in the simulation, our analysis predicts that 6 filaments would be sufficient. Simulations at these and other values of $\kappa$ and $\gamma$ suggest that filament bundle stiffnesses of roughly $1.5$ times the analytical values are required for filopodium stability, which seems to be due to the ability of the membrane to locally adapt to the helicity of the membrane, lowering the free energy barrier to collapse.

This, however, does not change the argument of the previous section: filopodium collapse can, for any reasonable $\kappa$ and $\gamma$, always be overcome with a finite --- and small --- number of filaments in the filament bundle.

\section{Conclusion}

Our simulations, combined with the observations of localized buckling in Ref.~\cite{Kuchnir1997} and the observation of long filopodia in systems without actin filament bundling proteins\cite{Liu}, suggest that filopodia with relatively small numbers of un-cross-linked filaments can be stable against classical Euler buckling. The filament bundles inside the filopodia are predicted to adopt a  helically buckled conformation, in accordance with the energetic considerations of section~\ref{sec:finrad}. In this conformation, the filament can still continue to grow. 

Although there is experimental evidence for the stability of filopodia to buckling, observing the helically buckled filament bundle experimentally might prove challenging: the radius of the helix is only large enough to be resolved optically in the most marginally stable filopodia. Invasive visualization techniques, such as electron microscopy of fixed samples, could jeopardize the mechanical integrity of membrane or filaments.

It should be noted that the mechanisms leading to helical buckling are not necessarily restricted to filopodia: this might happen in any comparable situation where a membrane exerts forces on a stiff filament or filament bundle, such as in cilia. Helically arranged filaments have, for example, been observed in non-spherical bacteria\cite{Young2003}. The mechanism described here might be able to account for both the symmetry breaking and the helical pitch in such arrangements.

We thank David Richmond, Allen Liu, as well as Steve Whitelam, Lutz Maibaum, Joshua Shaevitz, and Gerbrand Koster for helpful discussions. This work is supported in part by the California Institute for Quantitative Biosciences, by the National Science Foundation, and the National Institutes for Health Nanomedicine Center.

\bibliography{../bib}

\end{document}